**Letter**

# Heralded single photon source at 1550 nm from pulsed parametric down conversion

ALEXANDRE SOUJAEFF†*, SHIGEKI TAKEUCHI†, KEIJI SASAKI†, TOSHIO HASEGAWA‡, MITSURU MATSUI‡

†Japan Science and Technology Corporation-CREST Project; Research Institute for Electronic Science, Hokkaido University, Kita-12 Nishi-6, Kita-ku, Sapporo 060-0812, Japan, Tel +81-11-706-2648, Fax +81-11-706-3693
‡Mitsubishi Electric Corporation, Ofuna 5-1-1, Kamakura, 247-8501, Japan, TEL +81-467-41-2561, FAX +81-467-41-2185

*Corresponding author. Email: alex@es.hokudai.ac.jp

Heralding of single photon at 1550 nm from pump pulsed non degenerate spontaneous parametric downconversion is demonstrated. P(1) and P(2) of our source are 0.1871 and $2.4 \times 10^{-3}$ respectively. Triggering of our source is $2.16 \times 10^5$ trigger.s$^{-1}$. This source may be used in QKD system.

**1 Introduction**

Single Photon Source (SPS) is a basic element in many experiments in quantum information technology. Quantum Key Distribution (QKD) using BB84 protocol relies on the creation of pure single photon state as a carrier for information. However, the attenuated pulsed laser diode used actually for QKD emits pulses with multiple photons with a certain probability. As pointed in reference [1], the existence of such multi-photon pulses allows an eavesdropper to use powerful attacks in the presence of loss on the channel, and seriously limits the distance over which secure transmission is achievable. Development of SPS for QKD is important to extend the security range of current QKD systems, and many different approaches are investigated to realize on demand single photon states [2-3]. Heralded SPS is actually one of the most promising schemes. Such a source was realized at 1550 nm using CW pumping [4-5] and recently highly efficient entangled photon pair generation with CW pump using poled material was reported [6]. However, with CW pumping the heralded photons are created at random in time and it causes difficulty in synchronizing the sender and the remote receiver. On the other hand, heralded source with pulsed laser pumping, which produces single photon pulses with accurate timing, has only been realized [7] at 780 nm but not at telecom wavelength.

In this letter we present to our knowledge the first heralded SPS from pulsed Spontaneous Parametric Down Conversion (SPDC) at 1550 nm. The single photons are created in a very narrow time window defined by the pump and coupled to Single Mode Fiber (SMF) making this source directly usable for QKD. P(1) of our source is 0.1871, P(2) is $2.4 \times 10^{-3}$ and the actual source rate is $2.16 \times 10^5$ trigger.s$^{-1}$.

## 2 Principle

In the process of SPDC a photon from the pump incident on a non linear crystal is converted into two photons of lower energy if energy and momentum conservation are fulfilled through phase matching in the crystal. As proposed originally by Hong and Mandel [8] this phenomena can be used to create single photon state by detection of one photon of the pair heralding the presence of the other photon. In our case, phase matching condition has to be chosen according to the characteristics of QKD systems and current technical limitations of apparatus like single photon detectors [1]. Heralded photon should be produced at 1550 nm where losses are minimal in SMF. Photon of the pair to be detected and serve as a heralding signal should be produced in the visible region of the optical spectrum where commercially available silicon detectors offer high quantum efficiency and low noise in a free running operation mode.

We choose a BBO crystal with a cut angle of 26.42 degree pumped at the wavelength of 390 nm. For type I non linear interaction, signal and idler photons are created at 521 nm and 1550 nm collinearly with the pump beam. Following Kurtsiefer [9] recommendation for SPDC coupling, we defined target modes that represent the emitted down-converted photons using Gaussian optic. The modes can be effectively coupled into SMF in idler and signal paths.

## 3 Experimental set up

Experimental set up is shown in figure1. A mode-locked Ti:Sapphire laser (Tsunami, Spectra Physics) associated with a second harmonic generation unit produces 150 fs pulses at 390 nm (2.4 nm FWHM) at a rate of 82 MHz. A Pellin-Broca prism removes remaining infrared radiation from the pump laser. The beam is focused into the crystal (5 mm length with a plano-convex lens ($f$ = 700 mm) to a beam waist of 80 μm. Signal and idler photons are separated using a dichroic mirror oriented at 45 degree. Signal is reflected (R=99%) and idler is transmitted (T=97%) and both are then collimated by bi-convex lenses of 220 mm and 70 mm focal length respectively. In the signal path a dielectric longpass filter oriented at 45 degree and a bandpass filter (center wavelength 521 nm, 10 nm FWHM) reject residual pump. Total optical transmission coefficient is

46.6% between crystal and Single Photon Counting Module (SPCM) in our experiment. In idler path, a longpass colored glass filter is inserted. Photons are finally focused on fiber tip using aspheric lens of 11 mm focal length in both path. Total optical transmission coefficient is 81.7% between crystal and fiber. We use standard SMF at 520 nm and 1550 nm. Analyzing the influence of this optical system on the parametric fluorescence [7] using tuning curve [10] showed that parametric fluorescence is coupled over 6 nm (FWHM) at 520 nm, while corresponding heralded photon coupled into fiber are spread over 18 nm.

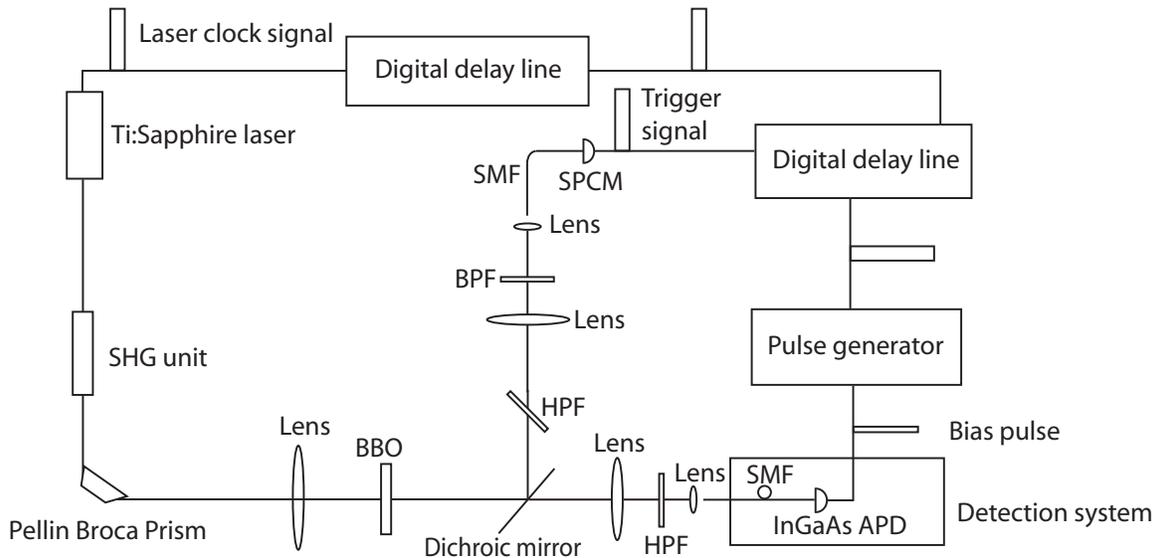

Figure 1. Schematic of the experimental set up: BPF, BandPass Filter; HPF, HighPass Filter.

Detection of signal is done with a SPCM (Perkin Elmer, SPCM-AQR-14 FC) fiber connected (54,7 % quantum efficiency at 520 nm, dark count 90 cps). The output signal from the SPCM is sent to a digital delay line (DG535, Stanford Research system) which provides the heralding electric signal, used as a trigger for our custom made 1550 nm detector consisting of an avalanche photodiode (Epitaxx EPM239) cooled down and operated in gated Geiger mode. Because of electronic delays in the trigger system, a SMF with a length of 55 m is inserted in the idler path after fiber coupler to allow synchronization of the 1550 nm APD. This induces additional optical losses in the idler path (76.5% transmission). For single count experiments at 1550 nm, this detector can also be triggered by the external clock which is generated from the clock signal of the pumping laser ( 82 MHz ) using an another digital delay line. Quantum efficiency, noise and afterpulsing probability (at 200 KHz) of our detector are respectively 10%, $2.5 \times 10^{-4}$ count per gate and 0.1%. Detection signals are counted with a photon counter (SR400, Stanford Research System).

## 4 Experimental results

We first measured the bandwidth of signal photon coupled into fiber with a spectrometer. FWHM bandwidth is 5 nm for a center wavelength of 521 nm, close to our prediction made using tuning curve (6 nm). Then, we measured photon emission in signal and idler path after fiber coupling. Single counts values were $2.90 \times 10^5$ cps and 285 cps respectively for 240 mW of pump power. The difference between these two values is due to the low triggering rate at idler wavelength (205 KHz) compared to signal. Coincidence count measurement for a trigger rate of $2.16 \times 10^5$ trigger.s$^{-1}$ was 3053 cps.

In order to find P(0), P(1) and P(2), the probability to have 0, 1 or 2 photons present at the output of the source at 1550 nm, we first measured the statistics of fiber coupled photon in the signal path with a Hanbury Brown and Twiss set up. The signal photons coupled into SMF were sent on a 50/50 bulk beam splitter and then coupled into multimode fiber and detected by a SPCM in each path of the beam splitter. The second order correlation function, $g^{(2)}(0)$ of signal photon coupled into fiber was found to be 0.95±0.5, characteristic of the poissonian statistics of coherent state as expected theoretically [11].

**5 Photon number distribution analysis**

From single count measurements in idler and signal path and also coincidence counts measurements, it is possible to calculate the probability distribution of the heralded photons under the assumption that the average number of photon pairs created in one pulse is low. For detailed analysis of the method, the reader should refer to reference [12]. The total number of pair created per second at the crystal output is estimated to be $6.8 \times 10^6$ photon/s (average photon number 0.0829). Mode coupling coefficient for idler and signal were found to be 22.00% and 16.87% respectively. These coefficients represent the portion of emitted photon effectively collected into SMF, and do not include optical losses. We obtained the probability P(0)=0.8096, P(1)=0.1871 and $P(2)=2.4 \times 10^{-3}$ for our source. If we compare it to the one of an attenuated coherent source with same P(1), P(2) of our source is 9 times lower.

P(1) is limited by the non unity transmission in the idler path before the fiber and mode coupling efficiency. We see no reasons not to obtain higher P(1) through optimization of our optical set up, possibly as pointed in a recent work [13] using numerical model to optimize coupling. P(2) can be tuned for a specific QKD system to allow secure transmission over a given distance adjusting pump power at the detriment of source rate [1]. Source rate depends on pump power, and is limited by the dead time of the digital delay line in the trigger circuit (1 µs). If two subsequent triggers from the SPCM are separated by less than 1 µs, the trigger signal is lost. This explains the difference between single count rate at 520 nm and rate of the source. Note that it doesn't degrade the heralding process. This can be solved if we use electronics with higher bandwidth.

**6 Conclusion**

In summary, we have demonstrated the first pulsed heralded source of single photon at 1550 nm. With a rate of $2.16 \times 10^5$ trigger.s$^{-1}$, P(1) of 0.1871, P(2) of $2.4 \times 10^{-3}$, our source may be usable for QKD experiment. Further experiments include photon bunching experiment, and P(2) reduction using photon resolving detector like visible light photon counter [14-15].

Acknowledgements:

The authors thank H. Fujiwara, R. Okamoto, and T. Tsurumaru for fruitful discussions and T. Chiba for experimental help. This work was supported by the Japan Science and Technology –Core Research Program for Evolution Science and Technology project, Japan and Mitsubishi Electric Corporation.